\documentclass[%
 reprint,
 amsmath,amssymb,
 aps,
]{revtex4-2}

\usepackage{graphicx}
\usepackage{dcolumn}
\usepackage{bm}

\usepackage{graphicx}
\usepackage{color}
\usepackage{bm}
\usepackage{epsfig}
\usepackage{epstopdf}
\usepackage{lineno,hyperref}

\def \b{\beta}
\def \l{\lambda}

\def \be{\begin{equation}}
\def \ee{\end{equation}}
\def \ben{\begin{eqnarray}}
\def \een{\end{eqnarray}}

\def \t{\theta}
\def \P{\Phi}

\def \m{\mu}
\def \x{\xi}

\def \La{\mathcal{L}}

\begin{document}


\title{Null geodesic structure for the Barriola-Vilenkin spacetime via {\bf k-}essence}

\author{Bivash Majumder}
 \email{bivashmajumder@gmail.com}
\affiliation{\textit{\textit{Department of Mathematics, Prabhat Kumar College\\
Contai, Purba Medinipur 721404, India}}
}%

\author{Saibal Ray}
 \email{saibal.ray@gla.ac.in}
\affiliation{\textit{Centre for Cosmology, Astrophysics and Space Science (CCASS), GLA University\\ Mathura 281406, Uttar Pradesh, India\\Institute of Astronomy Space and Earth Science\\Kolkata 700054, India}
}%

\author{Goutam Manna$^a$}
 \email{goutammanna.pkc@gmail.com}
\affiliation{\textit{Department of Physics, Prabhat Kumar College\\ Contai, Purba Medinipur 721404, India\\Institute of Astronomy Space and Earth Science\\Kolkata 700054, India}\\$^a$Corresponding author}


\begin{abstract}
Based on the work of Chandrasekhar [{\it The Mathematical Theory of Black Holes, Oxford Univ. Press (1992)}], we investigate the null geodesic structure of the emergent Barriola-Vilenkin spacetime in the context of {\bf k-}essence theory. For {\bf k-}essence, the emergent gravity metric is a one-to-one correspondence with the Barriola-Vilenkin (BV) metric connected to the Schwarzschild background, where the global monopole charge is replaced by the dark energy density. This equivalence holds specifically for a certain class of {\bf k-}essence scalar fields that have been constructed by Gangopadhyay and Manna [Euro. Phys. Lett., 100, 49001, (2012)]. We have traced out different trajectories for null geodesic in the presence of dark energy for the {\bf k-}essence emergent Barriola-Vilenkin spacetime. It is demonstrated that the outcomes deviate from the typical Schwarzchild spacetime owing to the fundamental configuration with a constant dark energy density. 

\end{abstract}
\keywords{K-essence geometry, Dark energy, Null Geodesic}

\maketitle

\section{Introduction}
Chandrasekhar~\cite{chandra} described the null geodesic structure of the Schwarzschild spacetime in his book on Black Holes. The author provides a detailed analysis of several types of orbits, including radial geodesic orbits, critical orbits, orbits of the first kind, orbits of the second kind and orbits with imaginary eccentricities. The accompanying pictures enhance the understanding of these concepts.

The geodesic structure of the Schwarzschild spacetime has also been addressed by Berti et al.~\cite{berti}. In the present context, Majumder and colleagues~\cite{bm1} have provided a comprehensive analysis of the temporal geodesic configuration pertaining to the emergent Barriola-Vilenkin spacetime within the framework of {\bf k-}essence. The article by Cruz et al.~\cite{cruz} gives an extensive review of the radial and non-radial trajectories associated with time-like and null geodesics structures in the Schwarzschild anti-de Sitter spacetime.

The time-like and null geodesic structures with figures for Bardeen spacetime have been shown in the ref.~\cite{zhou}. The solution of the Einstein equation governing the behaviour of the fields outside the core of a monopole has been derived by Barriola and Vilenkin in 1989~\cite{barriola}. The resulting black hole carries the global monopole charge where a global monopole falls under a Schwarzschild black hole. 

The following studies~\cite{babi1,babi2,babi3,babi4,babi5,scherrer1,scherrer2} offer an extensive investigation of the {\bf k-}essence theory, which proposes dominance of the kinetic energy over the potential energy. In the work by the authors~\cite{gm1}, it has been demonstrated that the emergent gravity metric, denoted by $\bar G_{\mu\nu}$, exhibits conformal equivalence to the Barriola-Vilenkin (BV) metric~\cite{barriola} within the Schwarzschild background. This equivalence is established for a specific configuration of the {\bf k-}essence scalar field $(\phi)$, which is based on the Dirac-Born-Infeld (DBI) model~\cite{born1,born2,born3}. Notably, in this configuration, the global monopole charge replaced by the constant kinetic energy ($\dot\phi^{2}=K$) of the {\bf k-}essence scalar field. They have chosen a form of the Lagrangian as $\La=-V(\phi)F(X)$ where $X=\frac{1}{2}g^{\mu\nu}\nabla_{\mu}\phi\nabla_{\nu}\phi$ with non-canonical kinetic terms. The dynamical solutions of the {\bf k-}essence equation of motion, which are not trivial, exhibit a spontaneous breaking of the Lorentz invariance and generate metric changes as a result of perturbations surrounding these solutions. This behavior distinguishes {\bf k-}essence from the relativistic field theories with canonical kinetic components. The metric employed for these perturbations in the {\it emergent or analogue} curved spacetime differs from the conventional gravitational metric ~\cite{babi1,babi2,babi3,babi4,babi5}.

In their recent publications, Manna et al.~\cite{gm2,gm3,gm4,gm5,gm6,gm7,gm8,gm9,gm10,gm11} have explored a range of topics, including the thermodynamics of black holes, gravitational collapse, the relationship between {\bf k-}essence and Vaidya geometry, features related to the Raycahudhuri equation and modified theory of gravity etc. These investigations have utilized the framework of {\bf k-}essence geometry, whch has been shown to be consistent with the observational findings~\cite{planck1,planck2}.

It is important to remember that the conventional (canonical or standard) theories do not provide a comprehensive understanding of every aspect of the physical scenario. The provided information lacks clarity in explaining the nature of dark matter, dark energy, the mechanisms behind the Big Bang, the disparity between matter and antimatter, the cosmological constant problem, the dimensions and configuration of the universe, cosmic inflation, the horizon problem, and other relevant aspects within the realm of cosmos. The primary unresolved issue in the field of basic physics is the reconciliation of gravity and quantum mechanics within a unified theoretical framework. So, in this direction, there have been a lot of works initiated by several scientists, and still work is to be done. 

In this context, we shall discuss the significance of non-standard theories such as the {\bf k-}essence theory \cite{yoo}. The {\bf k-}essence model is characterized by the presence of a non-canonical kinetic energy term in the Lagrangian, which has the ability to produce cosmic acceleration at late times without relying on potential energy. An attractor is defined as a point A that exerts an attractive force on nearby points, causing them to gravitate towards it.  The categorization of attractor solutions for {\bf k-}essence is divided into two distinct categories, as shown in previous studies \cite{picon1,picon2,kang}.
The ﬁrst one is tracker solution, in which {\bf k-}essence mimics the equation of state (EOS) of the background component in the Universe whereas in the second scenario, {\bf k-}essence exhibits a preference towards an EOS that deviates from that of matter or radiation. As a result of the attractor behavior, the {\bf k-}essence model is also insensitive to initial conditions. In contrast to the quintessence model, the {\bf k-}essence field exclusively tracks the radiation background, hence avoiding the need for fine-tuning that was present in the quintessence model. In addition, the coincidence problem is resolved by the presence of an S-attractor that attracts shortly after the beginning of the matter-dominated phase. However, the {\bf k-}essence framework does not explain why vacuum energy is so small. 

Now we are going to explore the significance of the non-canonical version of the Lagrangian within an alternative setting. The Lagrangian can be generally defined in either canonical or standard form as $L=T-V$, where $T$ represents the kinetic energy and $V$ represents the potential energy of the system. However, as indicated by Goldstein and Rana \cite{Goldstein,Rana}, the general form of the Lagrangian is non-canonical, whereas the canonical form is derived under specified conditions. The uniqueness of the functional form of $L$ is not guaranteed, as the Euler-Lagrange equations of motion can be satisfied by many Lagrangian choices \cite{Goldstein,Rana}. In addition, Raychaudhuri \cite{Raychaudhuri} highlights that when moving outside the realm of mechanics, the conventional notions of kinetic and potential energies become inappropriate. Consequently, the equation $L=T-V$ loses its applicability. Since we already have the field equations and need to determine a Lagrangian density to fix them, it may have started as a back calculation. Furthermore, it should be noted that the classical notion of $L (=T-V)$ is no longer valid within the framework of special relativistic dynamics. Consequently, it may be asserted that the general form of the Lagrangian is of a non-canonical nature \cite{gm9}.

In this work, our main motivation is thoroughly to investigate and trace out the null geodesic structures for the {\bf k-}essence emergent Barriola-Vilenkin (BV) type spacetime in the presence of dark energy based on the following work \cite{chandra}, however not in the context of Jacobi metric \cite{gibbons,chanda}. 

The paper is organized as follows:
In Section II, we have briefly reviewed the {\bf k-}essence geometry where the metric $\bar G_{\mu\nu}$ contains the dark energy field $\phi$ (i.e., the {\bf k-}essence scalar field) which satisfies the emergent gravity equations of motion. In Section III, we have discussed about the null geodesic structure for the {\bf k-}essence emergent Barriola-Vilenkin (BV) type spacetime in the presence of dark energy and also traced out the trajectories by considering dark energy density  in unit of critical density \cite{gm1,gm2,gm3} which is approximately $0.7$ \cite{planck3,planck4,planck5}. The conclusion of our work is in the last Section IV.

\section{{\bf k}-essence theory}
The action of the {\bf k-}essence geometry is given by \cite{babi1,babi2,babi3,babi4,babi5,scherrer1,scherrer2}
\ben
S_{k}[\phi,g_{\mu\nu}]= \int d^{4}x \sqrt{-g} \La(X,\phi),
\label{1}
\een
where $X={1\over 2}g^{\mu\nu}\nabla_{\mu}\phi\nabla_{\nu}\phi$ and the energy-momentum tensor is
\ben
T_{\mu\nu}\equiv {2\over \sqrt {-g}}{\delta S_{k}\over \delta g^{\mu\nu}}= \La_{X}\nabla_{\mu}\phi\nabla_{\nu}\phi - g_{\mu\nu}\La,
\label{2}
\een
where $\La_{\mathrm X}= {d\La\over dX},~~ \La_{\mathrm XX}= {d^{2}\La\over dX^{2}},
~~\La_{\mathrm\phi}={d\La\over d\phi}$ and  $\nabla_{\mu}$ is the covariant derivative defined with respect to the gravitational metric $g_{\mu\nu}$.

The scalar field equation of motion (EOM) is \cite{babi2,babi3,babi5}
\ben
-{1\over \sqrt {-g}}{\delta S_{k}\over \delta \phi}= \tilde G^{\mu\nu}\nabla_{\mu}\nabla_{\nu}\phi +2X\La_{X\phi}-\La_{\phi}=0,
\label{3}
\een
with 
\ben
\tilde G^{\mu\nu}\equiv \La_{X} g^{\mu\nu} + \La_{XX} \nabla ^{\mu}\phi\nabla^{\nu}\phi,
\label{4}
\een
and $1+ {2X  \La_{XX}\over \La_{X}} > 0$. Here $\La_{X}\neq 0$ for $c_{s}^{2}$ to be positive definite.

Using the conformal transformations
$G^{\mu\nu}\equiv {c_{s}\over \La_{x}^{2}}\tilde G^{\mu\nu}$ and $\bar G_{\mu\nu}\equiv {c_{s}\over \La_{X}}G_{\mu\nu}$, with
$c_s^{2}(X,\phi)\equiv{(1+2X{\La_{XX}\over \La_{X}})^{-1}}$ we have \cite{gm1,gm2,gm3,gm4,gm5,gm6}
\ben \bar G_{\mu\nu}
={g_{\mu\nu}-{{\La_{XX}}\over {\La_{X}+2X\La_{XX}}}\nabla_{\mu}\phi\nabla_{\nu}\phi}.
\label{5}
\een	

If $\La$ is not an explicit function of $\phi$ then the EOM (\ref{3}) is reduces to
\ben
-{1\over \sqrt {-g}}{\delta S_{k}\over \delta \phi}
= \bar G^{\mu\nu}\nabla_{\mu}\nabla_{\nu}\phi=0.
\label{6}
\een

It should be noted that in the case of non-trivial spacetime configurations of the field $\phi$, the resulting metric $\bar{G}_{\mu\nu}$ does not generally exhibit conformal equivalence to the metric $g_{\mu\nu}$. The scalar field $\phi$ exhibits distinct features compared to canonical scalar fields, and its local causal structure differs from those described by the metric tensor $g_{\mu\nu}$. 

We consider the DBI type Lagrangian as \cite{gm1,gm2,gm3,born1,born2,born3}
\ben
\La(X,\phi)= 1-V(\phi)\sqrt{1-2X},
\label{7}
\een
for $V(\phi)=V=\text{constant}$~\text{and}~$\text{kinetic~energy~of}~\phi>>V$, i.e., $(\dot\phi)^{2}>>V$. The presented Lagrangian form is commonly seen in the context of {\bf k}-essence fields, where the dominance of kinetic energy over potential energy is prominent. Then $c_{s}^{2}(X,\phi)=1-2X$. For scalar fields $\nabla_{\mu}\phi=\partial_{\mu}\phi$. Thus the effective metric (\ref{5}) is
\ben
\bar G_{\mu\nu}= g_{\mu\nu} - \partial _{\mu}\phi\partial_{\nu}\phi.
\label{8}
\een

The geodesic equation that corresponds to the {\bf k}-essence theory, expressed in terms of the new Christoffel connections denoted as $\bar\Gamma$, may be written as follows \cite{gm1,gm2,gm3}:
\ben
\frac {d^{2}x^{\alpha}}{d\l^{2}} +  \bar\Gamma ^{\alpha}_{\mu\nu}\frac {dx^{\mu}}{d\l}\frac {dx^{\nu}}{d\l}=0,
\label{9}
\een
where $\l$ is an affine parameter and  
\ben
\bar\Gamma ^{\alpha}_{\mu\nu} 
&=&\Gamma ^{\alpha}_{\mu\nu} -\frac {1}{2(1-2X)}\Big[\delta^{\alpha}_{\mu}\partial_{\nu}X
+ \delta^{\alpha}_{\nu}\partial_{\mu}X\Big].~~~~~~~~~~~
\label{10}
\een

In k-essence geometry, it is worth noting that Einstein's field equation can be expressed as
\ben
\mathcal{\bar{G}}_{\mu\nu}=\bar{R}_{\mu\nu}-\frac{1}{2}\bar{G}_{\mu\nu}\bar{R}=\kappa \bar{T}_{\mu\nu}, \label{eq:1}
\een
where $\kappa=8\pi G$ is constant, $\bar{R}_{\mu\nu}$ is Ricci tensor, $\bar{R}~ (=\bar{R}_{\mu\nu}\bar{G}^{\mu\nu})$ is the Ricci scalar and $\bar{T}_{\mu\nu}$ is the energy-momentum tensor of this geometry. The emergent energy-momentum tensor ($\bar{T}_{\mu\nu}$) can be determined to solve the left-hand side of the emergent Einstein field equation (\ref{eq:1}).

\section{Null Geodesics for the Barriola Vilenkin type emergent spacetime}
The authors \cite{gm1} have demonstrated that the emergent metric $\bar G_{\mu\nu}$ (\ref{8}) precisely correlated with the Barriola-Vilenkin (BV) metric, given a certain form of the {\bf k-}essence scalar field. This correlation is obtained when the standard gravitational metric $g_{\mu\nu}$ is assumed to be Schwarzschild.  The global monopole charge has been substituted with the constant kinetic energy of the scalar field.

The {\bf k-}essence emergent BV \cite{gm1}  type  metric is 

\begin{align}
  ds^{2} &=(1-{2GM\over r}-K)dt^{2}
-{1\over(1-{2GM\over r}-K)}dr^{2}\nonumber\\
 &-r^{2}d\theta^{2}-r^{2}sin^{2}\theta d\Phi^{2}\nonumber\\
 &= (\b-{2GM\over r})dt^{2}-\frac{dr^{2}}{(\b-{2GM\over r})}-r^{2}d\theta^{2}\nonumber\\
 & -r^{2}sin^{2}\theta d\Phi^{2},
\label{11}  
\end{align}

where $K$ is the constant kinetic energy (i.e., the dark energy density in unit of critical density \cite{gm1,gm2,gm3}) of the {\bf k-}essence scalar field and we define $\b=(1-K)$. The range for $K$ is $0<K<1$. The expression for the {\bf k-}essence scalar field is  $\phi(r,t)=\phi_{1}(r)+\phi_{2}(t) =\sqrt{K}[r+2GM~ln(r-2GM)]+\sqrt{K}t $. The kinetic part of this field is $\dot{\phi}^{2}\equiv \dot{\phi}_{2}^{2}=K$. If we consider $K=0$ then the {\bf k-}essence theory is meaningless and if $K=1$, then (\ref{11}) does not have a Newtonian limit \cite{schutz}. Also if we consider $K>1$ then the signature of the above metric (\ref{11}) is ill-defined and also we have the total energy density ($\Omega_{matter} + \Omega_{radiation} + \Omega_{dark~energy} = 1$) cannot exceed unity \cite{gm1}. It is also worth mentioning that the {\bf k-}essence emergent BV metric (\ref{11}) can be found by solving the emergent Einstein field equation (\ref{eq:1}).

The investigation of the geodesics equation in the emergent spacetime of the Barriola-Vilenkin type may be accomplished by deriving it from the following Lagrangian (as discussed in  \cite{chandra,cruz,berti,bm1}):

\begin{align}
 2\mathcal{L}&=(\b-{2GM\over r})\dot t^{2} -{1\over(\b-{2GM\over r})}\dot r^{2} \nonumber\\
 &-r^{2}\dot {\t}^{2}-(r^{2}sin^{2}\t) \dot {\P}^{2}
\label{12}   
\end{align}

where
$\dot t = \frac{dt}{d\tau},~\dot r = \frac{dr}{d\tau},~\dot \t = \frac{d\t}{d\tau},~\dot \P = \frac{d\P}{d\tau}$, $\tau$ is the proper time.

Therefore, momenta are
\begin{align}
  &p_{t}=(\b-{2GM\over r})\dot t,~p_{r}={1\over(\b-{2GM\over r})}\dot{r}, \nonumber\\
  &p_{\t}=r^{2}\dot{\t},~p_{\Phi}=r^{2}sin^{2}\t\dot{\P}
\label{13}  
\end{align}

and the Hamiltonian is
\ben
\mathcal{H}&&=p_{\m}\dot x^{\m}-\mathcal{L} \nonumber\\
&&=p_{t}\dot{t}-(p_{r}\dot{r}+p_{\t}\dot{\t}+p_{\P}\dot{\P})-\mathcal{L} \nonumber\\
&&=\mathcal{L}.
\label{14}
\een

For spherically symmetric nature of the metric, here the Lagrangian does not depend on $t$ and $\Phi$.  So  the equations of motions are $\dot{p}_{t}=0,~\dot{p}_{\P}=0$ which implies that \cite{bm1} 
\ben
&&\Big(\b-{2GM\over r}\Big)\dot{t}=\text{constant}=E \text{(say)},  \label{15} \\
&&r^{2}sin^{2}\t\dot{\P}=\text{constant}, \label{16}\\
&&\frac{d}{d\tau}\Big(r^{2}\dot{\t}\Big)=(r^{2}\sin\t\cos\t)\dot{\P}^{2}.
\label{17} 
\een

In \cite{bm1}, they also have considered that there is no contribution of potential energy since in the {\bf k-}essence theory, the contribution of the kinetic energy part dominates over the potential energy i.e., $K.E.>>P.E.$. Now to simplify, we consider the motion in the equatorial plane $\t=\frac{\pi}{2}$. Using the Eq. (\ref{16}), we have
\ben 
p_{\P}=r^{2}\dot{\P}=\text{constant}=L~\text{(say)},
\label{18}
\een
where $L$ is the angular momentum about an axis normal to the invariant plane. Finally using the Eqs. (\ref{15}), (\ref{18}) and (\ref{12})  the Lagrangian becomes
 \ben
2\mathcal{L}=\frac{E^{2}}{\b-\frac{2M}{r}}-\frac{\dot{r}^{2}}{\b-\frac{2M}{r}}-\frac{L}{r^{2}}.
\label{19}
\een

Now for time-like geodesic we consider $2\mathcal{L}=+1$  and for null geodesics $2\mathcal{L}=0$ . Here we only concentrate on the null geodesics.

\subsection{Null Geodesic}
Putting $\mathcal{L}=0$ in Eq. (\ref{19}), we get
\ben
\Big( \frac{dr}{d\tau} \Big)^{2}+ \frac{L^{2}}{r^{2}}\Big( \beta-\frac{2M}{r}\Big)=E^{2},
\label{eq:9}
\een
 with
 \ben
 \Big(\beta-\frac{2M}{r} \Big)\frac{dt}{d\tau}=E~\text{and}~ \frac{d\P}{d\tau}=\frac{L}{r^{2}},
 \label{eq:10}
 \een
by using Eqs. (\ref{15}) and (\ref{18}). Now substituting $r=\frac{1}{u}$ in Eq. (\ref{eq:9}), we obtain
\ben
\Big( \frac{du}{d\P}\Big)^{2}= 2Mu^{3}-\beta u^{2}+\frac{1}{D^{2}}= f(u)~\text{(say)},
   \label{eq:11}
\een
 and $D=\frac{L}{E}$ (say) which denotes the impact parameter. 
  
 To find different orbits, let us start with the equation 
 \ben
  f(u)=2Mu^{3}-u^{2}+\frac{1}{D^{2}}=0.
  \label{eq:17}
 \een

 Let the roots of the above equation are $u_{1}$, $u_{2}$ and $u_{3}$ then we have 
 \ben
&&u_{1}+u_{2}+u_{3}=\frac{\beta}{2M}, \label{eq:18a} \\
&&u_{1}u_{2}+u_{2}u_{3}+u_{1}u_{3}=0, \label{eq:18b} \\
&&u_{1}u_{2}u_{3}=-\frac{1}{2MD^{2}}.
\label{eq:18c}
 \een

 Now using Descartes' Rule of Sign, the Eq. (\ref{eq:17}) must have one negative root, since $M>0$ and $D^{2}>0$ and the other two roots are real or complex. For simplicity we assume that $u_{1}<0$. In this situation one may note that there arise three different cases, which are as follows:\\
 
 {\it Case-A:} One root is negative and the others two roots are positive and equal, i.e., $u_{2}=u_{3}$. In this case we have two types of  orbits: one is in the interval $0<u\le u_{2}$ and the other one is in the interval $u_{2}\le u< \infty$. In the first case the orbits are arriving from infinity and approaching the circle $r=\frac{1}{u_{2}}$ by spiralling around it whereas in the second case the orbits are starting from the aphelion distance $r=\frac{1}{u_{2}}$ and plunging to the singularity $r=0$. We shall term these two types of orbits as the orbits of the first and second kind, respectively. 

 {\it Case-B:} One root is negative and the others two roots are positive and distinct, i.e., $u_{2}<u_{3}$. Again in this case we have two kinds of orbits: the orbit of first kind is in the interval $0<u\le u_{2}$ and the orbit second kind is in the range $u_{3}\le u< \infty$.

 {\it Case-C:} One root is negative and the others two roots are complex. In this case there is no real roots, so the only possibility of tracing a orbit is in the interval $0<u<\infty$.  So these kinds of orbits are arriving from infinity and plunging to the singularity.
 
\subsubsection{ {\bf Case-A : Critical Orbits}}
Here, we first consider the critical case that is the first case where one root is negative and the others two roots are positive and equal. So for the occurrence of two coincident roots, we must have $$\frac{df(u)}{du}=0 \Rightarrow u=\frac{1}{3\bar{M}}$$
 where we define $\bar{M}=\frac{M}{\b}=\frac{M}{1-K}$, so that $\bar{M}>M$ as $K<1$ and neglecting the solution $u=0$.
 
 Therefore, we have
 \ben
u_{2}=u_{3}=\frac{1}{3\bar{M}}.
\label{eq:19a}
 \een
 
 Now $u_{2}$ and $u_{3}$ are the roots of the Eq. (\ref{eq:17}), then we have
 \ben
D=\frac{3\sqrt{3}\bar{M}}{\sqrt{\beta}}.
\label{eq:19} 
 \een

Here we can see that the impact parameter $D$ for null geodesics in the presence of dark energy density is much higher than the impact parameter for null geodesics in the usual Schwarzschild background \cite{chandra}. 

Now substituting Eq. (\ref{eq:19a}) in Eq. (\ref{eq:18a}), we get 
 \ben
u_{1}= -\frac{1}{6\bar{M}},~~~ u_{2}=u_{3}=\frac{1}{3\bar{M}}~\text{when}~D=\frac{3\sqrt{3}\bar{M}}{\sqrt{\beta}}
\label{eq:20}
 \een
 
Therefore from Eq. (\ref{eq:11})
\ben
\frac{du}{d\P}=-\sqrt{2M (u+\frac{1}{6\bar{M}})}\Big(u-\frac{1}{3\bar{M}}\Big).
\label{eq:21}
\een

Here we take the negative sign in the RHS, so that $\P$ may increase. Now solving the above Eq. (\ref{eq:21}), we get
\ben
u=-\frac{1}{6\bar{M}}+\frac{1}{2\bar{M}}\tanh^{2}{\frac{\sqrt{\beta}}{2}(\P-\P_{0})},
\label{eq:22}
\een
where $\P_{0}$ is a constant of integration. Again, $\P_{0}$ can be considered such a way that $$\tanh^{2}{\frac{\P_{0}\sqrt{\beta}}{2}}=\frac{1}{3}$$
 then
 \ben
&&u=0~\text{when}~\P=0 \nonumber\\
\text{and}~&&u=\frac{1}{3\bar{M}}~\text{when}~\P\rightarrow\infty.
\label{eq:23}
 \een
 
\begin{figure}[h]
  \centering
\includegraphics[scale=1.5]{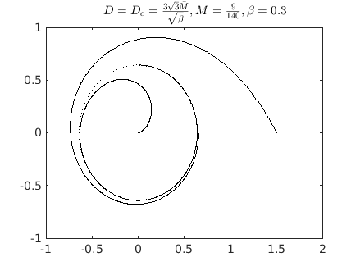}
 \caption{ The critical orbits of the first and second kind for the impact parameter $D=D_{c}=\frac{3\sqrt{3}\bar{M}}{\sqrt{\beta}}$ and for both cases $M=\frac{9}{140}$ and $\beta=0.3$.}
 \label{fig:1}
\end{figure}

 Thus the orbits of first kind with the impact parameter $D=\frac{3\sqrt{3}}{\sqrt{\beta}}\bar{M}$ arrives from infinity and asymptotically approaches the circle $r=3\bar{M}$ as shown in Fig. \ref{fig:1}. 
 
 Again, for the orbits of second kind, we substitute
 \ben
  u=\frac{1}{3\bar{M}}\frac{1}{2\bar{M}}\tan^{2}{\frac{\xi}{2}},
  \label{eq:24a}
 \een
 in Eq. (\ref{eq:21}), we get 
 \ben
\frac{d\xi}{d\P}= \sqrt{\beta} \sin {\frac{\xi}{2}}.
\label{eq:24}
 \een

In the above equation we consider the positive sign in the RHS so that we may $\P$ increase. Now integrating Eq. (\ref{eq:24}) we get
  \ben
\P=2\sqrt{\beta}\log{\tan{\frac{\xi}{4}}}, 
\label{eq:25}
  \een
and therefore using the Eqs. (\ref{eq:24a}) and (\ref{eq:24})
\ben
u=\frac{1}{3\bar{M}}+\frac{1}{\bar{M}}\frac{2e^{\frac{\P}{\sqrt{\beta}}}}{\Big(e^{\frac{\P}{\sqrt{\beta}}}-1\Big)^{2}}, 
\label{eq:26}
\een
which gives 
\ben
&&u\rightarrow\infty~\text{when}~\P\rightarrow 0 \nonumber\\
\text{and}~&&u=\frac{1}{3\bar{M}}~\text{when}~\P\rightarrow\infty.
\een

 Therefore starting the aphelion distance $r=3\bar{M}$, these orbits of second kind plunges to singularity ($r=0$) as shown in Fig. \ref{fig:1} where $M=\frac{9}{140}$ and $\beta=0.3$.
 
At this juncture let us discuss about the {\it cone of avoidance} which at any point can be defined by the solution of Eq. (\ref{eq:22}) whose generators are null rays which passing through that point. We will establish in Case-C that the light rays included in the cone are getting trapped after crossing the horizon. Let $\Psi$ be the half angle of the cone then
\ben
\cot{\Psi}=\frac{1}{r}\frac{dr^{'}}{d\P},
\label{eq:27}
\een
 where 
 \ben
 dr^{'}=\frac{1}{\sqrt{\beta-\frac{2M}{r}}}dr,
\label{eq:28}
 \een
where $dr^{'}$ denotes an element of the proper length along the generators of the cone.

Using Eqs. (\ref{eq:27}) and (\ref{eq:28}), we get
 \ben
\frac{du}{d\P}=-u\sqrt{\beta}\cot{\Psi}\sqrt{1-2\bar{M}u}.
\label{eq:29}
 \een

Again substituting Eq. (\ref{eq:29}) in Eq. (\ref{eq:21}), we get 
\ben
\cot{\Psi}= \frac{\Big(\frac{r}{3\bar{M}}-1\Big)\sqrt{\Big(\frac{r}{6\bar{M}}+1\Big)}}{\sqrt{\Big(\frac{r}{2\bar{M}}-1\Big)}},
\label{eq:30}
\een
which gives  
\ben
&&\Psi \sim \frac{3\sqrt{3}}{r}~\text{when}~r\rightarrow \infty, \nonumber\\
&&\Psi=\frac{\pi}{2}~\text{when}~r=3\bar{M}, \nonumber\\
     ~\text{and}~
&&\Psi=0~\text{when}~r=2\bar{M}.
\label{eq:31}
\een

\begin{figure}[h]
\centering
\includegraphics[scale=0.45]{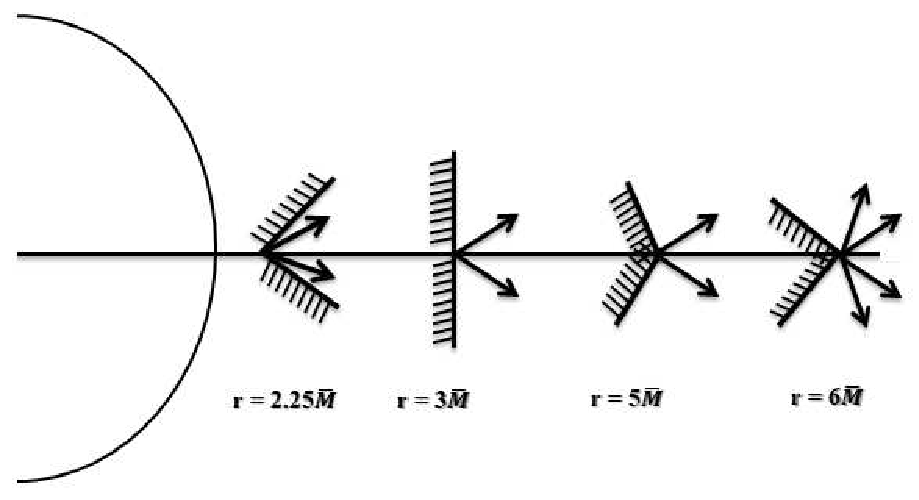}
\caption{Cone of Avoidance at various distances}
\label{fig:2}
\end{figure}

 Therefore the cone of avoidance become narrower directed inward when $r>3\bar{M}$ and the cone spread out fully at $r=3\bar{M}$ and directed outward when $r<3\bar{M}$ and again narrower when $r\rightarrow 2\bar{M}$ (see Fig. \ref{fig:2}). This figure is quite different in the presence of dark energy density $K=0.7$ \cite{planck1,planck2,planck3,planck4} from usual Schwarzschild Spacetime \cite{chandra} where the respective three case arise when $r>3M$, $r=3M$ and $r<3M$ since $\bar{M}>M$ as $\beta=0.3$.

\subsubsection{{\bf Case-B : Orbits of the First and Second Kind}}
Here we consider the second case where the one root is negative and the others two roots are positive and distinct. Let us assume that $u_{1}<0$, $0<u_{2}<u_{3}$.

So, let roots of the Eq. (\ref{eq:17}) are in the following forms:
 \ben
&&u_{1}=\frac{P-2\bar{M}-Q}{2\bar{M}P}, \nonumber\\
&&u_{2}= \frac{1}{P}, \nonumber\\
&&u_{3}=\frac{P-2\bar{M}+Q}{4\bar{M}P}.
\label{eq:32}
\een
 where $P$ is taken as the perihelion distance and $Q$ is a constant. 
 
 Now clearly it satisfies the Eq. (\ref{eq:18a}). To satisfy the Eqs. (\ref{eq:18b}) and (\ref{eq:18c}), we must have 
 \ben
&&Q^{2}=\Big(P-2\bar{M} \Big)\Big(P+6\bar{M} \Big), \label{eq:33a}\\
&&D^{2}=\frac{8\bar{M}P^{3}}{\beta \Big[Q^{2}-(P-2\bar{M})^{2}\Big]}.
\label{eq:33b}
  \een

To satisfy our assumption that $u_{2}<u_{3}$, we must have 
 \ben
P+Q-6\bar{M}>0,
\label{eq:33c}
 \een
 and for $u_{1}<u_{2}$, 
 \ben
P-Q-6\bar{M}<0.
\label{eq:33d}
 \een

 Now from the Eqs. (\ref{eq:33a}) and (\ref{eq:33b}), we get
  \ben
D^{2}=\frac{P^{3}}{\beta(P-2\bar{M})},
\label{eq:34a}
  \een
and from the Eqs. (\ref{eq:33a}) and (\ref{eq:33c})
 \ben
(P+6\bar{M})(P-2\bar{M})>(P-6\bar{M})^{2},
\label{eq:34b}
  \een

Simplifying, we get
  \ben
    P>3\bar{M}.
\label{eq:34c}
  \een

So, from the Eq. (\ref{eq:34a}), we have 
\ben
D>\frac{3\sqrt{3}\bar{M}}{\sqrt{\beta}}=D_{c}~\text{(say)}.
   \label{eq:34d}
\een
 
 Therefore, we can say that these kind of orbits are found when the impact parameter is greater than $\frac{3\sqrt{3}\bar{M}}{\sqrt{\beta}}$.

Again, from  Eqs. (\ref{eq:17}) and (\ref{eq:32}), we write
\ben
f(u)=2M\Big(u-u_{1}\Big)\Big(u-u_{2}\Big)\Big(u-u_{3}\Big).
\label{eq:35}
\een

 Substituting
 \ben
u=\frac{1}{P}-\frac{Q-P+6\bar{M}}{8\bar{M}P}(1+\cos{\xi})
\label{eq:35b}
 \een
 in Eq. (\ref{eq:35}) and using Eq. (\ref{eq:11}), we get
 \ben
\frac{d\xi}{d\P}=\sqrt{\Big(\frac{Q\beta}{P}\Big)}\sqrt{1-k^{2}\sin^{2}{\frac{\xi}{2}}},
\label{eq:35c}
 \een
 where $k^{2}=\frac{Q-P-6\bar{M}}{2Q}$.
 
In this above equation we consider the positive sign to keep $\P$ increasing. Now integrating the equation (\ref{eq:35c}), we get
\ben
\P=2\sqrt{\Big(\frac{P}{Q\beta}\Big)}\Big[\mathbf{K}(k)-\mathbf{F}\Big(\frac{\xi}{2},k\Big)\Big],
\label{eq:35a}
\een 
where $\mathbf{F}\Big(\frac{\xi}{2},k\Big)$ is the incomplete elliptic integral of the first kind and $\mathbf{K}(k)$ is the complete elliptic integral of the first kind. Therefore
\ben
&&\mathbf{K}(k)=\int^{\frac{\pi}{2}}_{0}\frac{dz}{\sqrt{1-k^{2}\sin^{2}{z}}}, \nonumber\\
&&\mathbf{F}\Big(\frac{\xi}{2},k\Big)=\int^{\frac{\xi}{2}}_{0}\frac{dz}{\sqrt{1-k^{2}\sin^{2}{z}}},
\label{eq:36}
\een
with $\frac{\xi}{2}=z$.

Thus we have
\ben
&&u=\frac{1}{P}~\text{and}~\P=0~\text{when}~\xi=\pi, \nonumber\\
&&u\rightarrow0 ~\text{and}~\P\rightarrow\P_{\infty}~\text{when}~\xi=\xi_{\infty},
\label{eq:37}
\een
where 
\ben
&&\P_{\infty}=2\sqrt{\Big(\frac{P}{Q\beta}\Big)}\Big[\mathbf{K}(k)-\mathbf{F}\Big(\frac{\xi_{\infty}}{2},k\Big)\Big], \nonumber\\\text{and}~&&\sin^{2}{\frac{\xi_{\infty}}{2}}=\frac{Q-P+2\bar{M}}{Q-P+6\bar{M}}.
\label{eq:38}
\een

\begin{figure}[h]
 \centering
\includegraphics[scale=1.4]{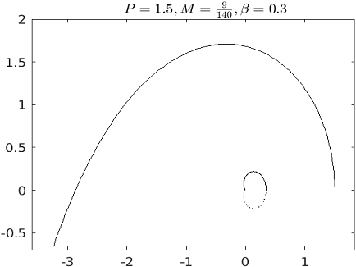}
 \caption{ The orbits of the first and second kind for $P=1.5$ and for both cases $M=\frac{9}{140}$ and $\beta=0.3$.}
 \label{fig:3}
\end{figure}

 Thus the range of $\xi$ is $\xi_{\infty}<\xi<\pi$. So we can conclude that starting from infinity (when $\P\rightarrow\P_{\infty}$), the orbits of first kind asymptotically approaches to $r=P$ (when $\P=0$) by spiralling around it. Now again in the presence of dark energy density $K=0.7$, these ranges of $\P$ and $r$ are different since $\bar{M}>M$ and similar in their orientation. In Fig. \ref{fig:3}, we have traced the orbits of first kind by considering $P=1.5$, $M=\frac{9}{140}$ with $\beta=0.3$. 
 
To obtain the orbits of the second kind, let us substitute 
\ben
u=\frac{1}{P}+\frac{Q+P-6\bar{M}}{4\bar{M}P}\sec^{2}{\frac{\chi}{2}}.
\label{eq:41}
\een

Then from the Eq. (\ref{eq:11}), we get 
\ben
\frac{d\chi}{d\P}=\sqrt{\frac{Q\beta}{P}\Big[1-k^{2}\sin^{2}{\frac{\chi}{2}}\Big]}.
\label{eq:42}
\een

In the above form we consider the positive sign in the RHS so that $\P$ may increase. Now integrating (\ref{eq:42}), one may get
\ben
\P=2\sqrt{\Big(\frac{P}{Q\beta}\Big)}\mathbf{F}\Big(\frac{\chi}{2},k\Big).
\label{eq:43}
\een
       
Therefore        
 \ben
&&u= \frac{Q+P-2\bar{M}}{4\bar{M}P}~\text{and}~\P=0~\text{when}~\chi=0, \nonumber\\
&&u\rightarrow\infty~\text{and}~\P=\mathbf{K}(k)~\text{when}~\chi=\pi.
\label{eq:44}
\een

 Here we have the range of $\chi$ is $0<\chi<\phi$. Thus starting from the aphelion distance $r=\frac{4\bar{M}P}{Q+P-2\bar{M}}$ when $\chi=\pi$, the orbit of second kind plunges to the singularity ($r=0$)  see Fig. \ref{fig:3} where we consider $M=\frac{9}{140}$ and $\beta=0.3$. Again the orientation of the orbits are similar with different values in the presence of dark energy density $K=0.7$ as we have seen in \cite{chandra}.

\subsubsection{{\bf Case-C : Orbits of with imaginary eccentricities and impact parameters less than $3\sqrt{3}\frac{\bar{M}}{\sqrt{\beta}}$}}

Finally, we are now discuss the nature of the orbits with imaginary eccentricities ($ie$) that is the  two roots of the Eq. (\ref{eq:17}) are imaginary and the other one is negative. To start with let us consider the roots of the Eq. (\ref{eq:17}) in the form 
\ben
u_{1}=\frac{1}{2\bar{M}}-\frac{2}{l},~u_{2}=\frac{1}{l}(1+\textit{i}e)~~and~~u_{3}=\frac{1}{l}(1-\textit{i}e).\nonumber\\
\label{eq:44a}
\een

Note that here we consider $e>0$, then from (\ref{eq:11}), we get
\begin{align}
& f(u)=2Mu^{3}-\beta u^{2}+2M\Big(\frac{e^{2}-3}{l^{2}}+\frac{1}{2\bar{M}l}\Big)u\nonumber\\
&-2M\frac{1+e^{2}}{l^{2}}\Big(\frac{1}{2\bar{M}}-\frac{2}{l}\Big).
    \label{eq:45}   
\end{align}

Now comparing this with the Eq. (\ref{eq:11}), we have
\ben
&&l-\bar{M}(3-e^{2})=0, \label{eq:46} \\
&&\frac{1}{2MD^{2}}=\Big(\frac{2}{l}-\frac{1}{2\bar{M}} \Big)~\frac{1+e^{2}}{l^{2}}.
\label{eq:45a}
\een

Taking
\ben
  \mu=\frac{M}{l\beta} \label{eq:49}  
\een
and then from Eq. (\ref{eq:46}), we get
\ben
    e^{2}=\frac{3\mu-1}{\mu} \label{eq:47}
\een
and from Eq. (\ref{eq:45a})
\ben
    \frac{D^{2}}{\bar{M}^{2}}=\frac{1}{\mu\beta(4\mu-1)^{2}}. \label{eq:48}
\een

Since $e^{2}>0$, we have
\ben
    \mu>\frac{1}{3} \label{eq:50}
\een
 and from the Eq. (\ref{eq:48})
 \ben
     D<\frac{3\sqrt{3}\bar{M}}{\sqrt{\beta}}. \label{eq:51} 
 \een

Thus to obtain these kind of orbits we must have the impact parameter $D$ must less than $\frac{3\sqrt{3}\bar{M}}{\sqrt{\beta}}$ and $\mu$ must greater than $\frac{1}{3}$. 

Now if we consider the substitution
\ben
  u=\frac{1}{l}\Big(1+e\tan{\frac{\xi}{2}}\Big)
  \label{eq:52}
\een
in Eq. (\ref{eq:11}), then we have
 \ben
\Big(\frac{d\xi}{d\P}\Big)^{2}=2\beta\Big[(6\mu-1)\cos{\xi}+2\mu e\sin{\xi}+(6\mu-1)\Big].\nonumber\\
     \label{eq:53}
 \een

 To simplify things we substitute 
 \ben
    \sin^{2}{\psi}=\frac{1}{\Delta+6\mu-1}\Big[\Delta-2\mu e\sin{\xi}-(6\mu-1)\cos{\xi}\Big] \nonumber\\ \label{eq:54} 
 \een
and then differentiating, we get
\ben
 \Big(\frac{d\psi}{d\xi}\Big)^{2}=\frac{\Delta}{2(\Delta+6\mu-1)\cos^{2}{\psi}}\Big[1-k^{2}\sin^{2}{\psi}\Big] \label{eq:55} 
\een
where
\ben
k^{2}=\frac{\Delta+6\mu-1}{2\Delta} \label{eq:56}    
\een
and 
\ben
\Delta^{2}=(6\mu-1)^{2}+4\mu^{2}e^{2}. \label{eq:57}    
\een

Combining Eqs. (\ref{eq:53}) and (\ref{eq:55})
\ben
\Big(\frac{d\P}{d\psi}\Big)^{2}=\frac{1}{\Delta\beta(1-k^{2}\sin^{2}{\psi})}. 
    \label{eq:58}
\een

 Integrating (\ref{eq:58}), we get
 \ben
\P=\frac{1}{\sqrt{\Delta\beta}}\Big[\mathbf{K}(k)-\mathbf{F}\Big(\psi,k\Big)\Big].
\label{eq:59}
 \een

  Therefore:
$$ when~~\xi=\pi,~u\rightarrow\infty~\text{and}~\psi=-\frac{\pi}{2},\frac{\pi}{2}$$
$$and~when~\xi\rightarrow\xi_{\infty},~u\rightarrow0~and~\psi\rightarrow\sin^{-1}{\frac{\Delta+1}{\Delta+6\mu-1}}$$
$$where~\xi_{\infty}=2\tan^{-1}{\frac{1}{e}}.
$$

\begin{figure}[h]
  \centering
\includegraphics[scale=1.5]{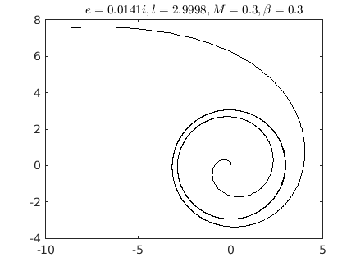}
\caption{The orbits of imaginary eccentricities with the impact parameter   $D<D_{c}=\frac{3\sqrt{3}\bar{M}}{\sqrt{\beta}}$ for $e=0.0141i$, $l=2.9998$ and $M=0.3$, $\beta=0.3$ }
 \label{fig:4}
\end{figure}

\begin{figure}[h]
 \centering
\includegraphics[scale=1.5]{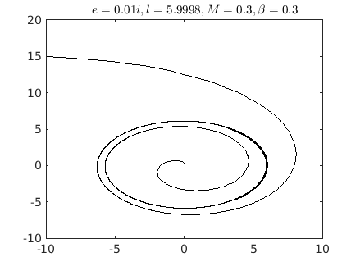}
 \caption{The orbits of imaginary eccentricities with the impact parameter   $D<D_{c}=\frac{3\sqrt{3}\bar{M}}{\sqrt{\beta}}$ for $e=0.01i$, $l=5.9998$  and $M=0.3$, $\beta=0.3$ }
 \label{fig:5}
\end{figure}

Therefore these kind of orbits are arriving from infinity when  $\xi\rightarrow\xi_{\infty}$ and plunging to singularity $r=0$ when $\xi=\pi$. In Figs. \ref{fig:4} and \ref{fig:5}, we have traced the trajectories with $e=0.0141i$, $l=2.9998$ and  $e=0.01i$, $l=5.9998$ respectively with $M=0.3$ and $\beta=0.3$.

\subsection{\textbf{Radial Geodesic}}
For radial geodesic the angular momentum is to be zero, therefore from the Eq. (\ref{eq:9}), we have
 \ben
\frac{dr}{d\tau}=\pm E
\label{eq:12}
 \een
and from the Eq. (\ref{eq:10})
\ben
&&\Big( \beta- \frac{2M}{r}\Big)\frac{dt}{d\tau}=E \nonumber \\
&&\Rightarrow t=\pm r^{*}+ \text{constant}_{\pm},
\label{eq:13}
\een
where 
\ben
r_{*}=\frac{1}{\beta}\Big[r+2\bar{M}log\Big(\frac{r}{2\bar{M}}-1\Big)\Big].
    \label{eq:14}
\een

Therefore 
\ben
\frac{d}{dr_{*}}=\frac{\Delta}{r^{2}}\frac{d}{dr},
    \label{eq:15}
\een
where $\Delta=\beta r^{2}\Big(1- \frac{2\bar{M}}{r}\Big)$ is the horizon function. 

Again, from Eq. (\ref{eq:14})
\ben
r\rightarrow 2\bar{M}+0\Rightarrow r_{*}\rightarrow -\infty\\~and~\rightarrow\infty\Rightarrow r_{*}\rightarrow\frac{r}{\beta},
\een
 and from Eq. (\ref{eq:12}), 
\ben 
 r=\pm E\tau +\text{constant}_{\pm}.
 \een
 
 This shows that the radial geodesic takes an infinite co-ordinate time to arrive at the horizon for an observer outside the horizon even though the radial geodesic crosses the horizon in its own proper time. This result is quite similar as in the Schwarzschild spacetime \cite{chandra} since $\beta$ is a constant.

\section{Conclusion}
All conceivable paths for null geodesics in the Barriola-Vilenkin spacetime arising from {\bf k-}essence have been systematically determined. According to the scholarly work authored by Chandrasekhar~\cite{chandra}, the presence of dark energy density significantly alters the ranges seen in Schwarzschild spacetime, as described in the book.

The determination of critical orbits is possible by identifying the condition under which the two roots of Eq. (\ref{eq:17}) are both positive and equal. In the case of critical orbits of the first kind, it has been shown that when initiated from an infinite distance, the orbits gradually approach the value of $r=3\bar{M}$ by spiralling around it. On the other hand, critical orbits of the second kind commence at $r=3\bar{M}$ and converge towards the singularity located at $r=0$. In both cases the radius of the circle $r=3\bar{M}$ are much higher than the than radius the circle $r=3M$ which can be seen in usual Schwarzschild spacetime. 
In our case the cone of avoidance, we found that that the cone opens out fully when $r=3\bar{M}$ which is higher than usual Schwarzschild spacetime where it opens out fully when $r=3M$.

By considering the two roots of the Eq. (\ref{eq:17}) are positive and distinct, we have traced the orbits of the first and second kind where we found that the orbits of the first kind are arriving from infinity and asymptotically approaches to $r=P$ (perihelion distance) and the orbits of the second kind are starting from $r=\frac{4\bar{M}P}{Q+P-2\bar{M}}$ and plunge to the singularity. Here also we note that these kinds of orbits can be found only when the impact parameter is greater than $D_{c}=\frac{3\sqrt{3}\bar{M}}{\sqrt{\beta}}$. It should be noted that the value of $D_{c}$ for this case is much much greater than the $D_{c}$ in the Schwarzschild Spacetime \cite{chandra}.

When the impact parameter is less than $D_{c}=\frac{3\sqrt{3}\bar{M}}{\sqrt{\beta}}$, the orbits of imaginary eccentricities can be traced, where arriving from infinity these kinds of the orbits are approaching to singularity $r=0$. 

For the case of radial geodesics, we have evaluated that these kinds of geodesics take infinite co-ordinate time to arrive at the horizon for an observer outside the horizon, although they cross the horizon in their own proper time, which is similar to Schwarzschild spacetime.

Finally, we conclude that although the orientation of trajectories are quite similar for both the Schwarzschild spacetime and the {\bf k-}essence emergent Barriola-Vilenkin spacetime but the ranges are much more different and the value of the $D_{c}$ is much higher due to the presence of dark energy density $K$.

\section*{Acknowledgments}

G.M. acknowledges the DSTB, Government of West Bengal, India for financial support through Grant Nos. 856(Sanc.)/STBT-11012(26)/6/2021-ST SEC dated 3rd November 2023.
S.R. is thankful to the Inter-University Centre for Astronomy and Astrophysics (IUCAA), Pune, India for providing Visiting Associateship
under which a part of this work was carried out who also gratefully
acknowledges the facilities under ICARD, Pune at CCASS, GLA University,
Mathura.




\end{document}